\documentstyle[version2,aps]{revtex}
\begin{document}
\draft
\columnsep -.375in
\twocolumn[
\preprint{P-95-04-020-i}
\begin{title}
Chaos in Andreev Billiards
\end{title}
\author{
Ioan Kosztin\rlap,$^{(a)}$
Dmitrii L.~Maslov$^{(a,b,d)}$ and
Paul M.~Goldbart$^{(a,b,c)}$
}
\begin{instit}
$^{(a)}$Department of Physics,
$^{(b)}$Materials Research Laboratory and
$^{(c)}$Beckman Institute, \\
University of Illinois at Urbana-Champaign,
Urbana, Illinois 61801, USA\\
$^{(d)}$Institute for Microelectronics Technology,
Academy of Sciences of Russia, Chernogolovka, 142432 Russia
\end{instit}
\begin{abstract}
A new type of classical billiard---the Andreev billiard---is
investigated using the tangent map technique. Andreev billiards consist
of a normal region surrounded by a superconducting region. In contrast
with previously studied billiards, Andreev billiards are integrable in
zero magnetic field, {\it regardless of their shape\/}. A magnetic
field renders chaotic motion in a generically shaped billiard, which is
demonstrated for the Bunimovich stadium by examination of both
Poincar\'e sections and Lyapunov exponents.  The issue of the
feasibility of certain experimental realizations is addressed.
\end{abstract}
\pacs{PACS numbers:
05.45.+b,
74.80.-g,
74.40.+k
\hfil\break
Preprint Number: P-95-04-020-i
}
]
\narrowtext
\newpage
In the development of the understanding of chaos, a prominent role has
been played by the class of classical mechanical systems known as
billiards \cite{REF:mvbejp}.  In such systems, particles are confined
by a step-like, single-particle potential to a region of space within
which they propagate ballistically.  Unless the shape of the billiard
is highly regular (e.g., circular), in which case the
system is integrable, the motion of a particle is chaotic.
Unpredictability, the hallmark of chaotic motion, can be diagnosed
qualitatively by the morphologies of Poincar{\'e} sections, and more
quantitatively by the corresponding Lyapunov exponents, the positivity
of which signals the exponential sensitivity of trajectories to initial
conditions.

A common feature of all versions of billiards studied to date
\cite{REF:mvbejp} is that reflection at boundaries is specular, i.e.,
only the component of the velocity {\it normal\/} to the boundary is
inverted. We refer to such versions as conventional billiards (CBs; see
Fig.~\ref{FIG:cartoon}a, left).  The purpose of this Letter is to
explore the issue of classical chaos in a novel class of billiards,
which have the property that scattering at boundaries is
retro-reflective, i.e., {\it all\/} components of the velocity are
inverted, as is depicted in Fig.~\ref{FIG:cartoon}a (right).  We refer
to such billiards as  Andreev billiards (ABs). Although we are unaware
of examples of such a reflection mechanism in the realm of classical
physics, a well-known example exists in condensed matter physics:
Andreev reflection of electronic quasiparticles (having energies in the
superconducting gap $\Delta$) from the normal-to-superconductor
interface \cite{REF:AFA,REF:quantum}.  Thus, we envisage ABs as normal
(N) domains surrounded by superconductor (S). It is adequate to regard
the motion of electronic quasiparticles as semiclassical, provided that
the billiard size is much larger than their  typical de~Broglie
wavelength.

The change from specular to Andreev reflection has a striking
consequence in the context of chaos:  whereas typical motion in a
generically-shaped CB is chaotic, motion in ABs
is integrable, regardless of the shape of the billiard.  This
integrability becomes evident from the observation
(cf.~Fig.~\ref{FIG:cartoon}a) that all motion occurs along chordal
trajectories connecting only two points on the boundary: what system
could be less ergodic?

The presence of a magnetic field ${\bf B}$ substantially alters the
situation (see Fig.~\ref{FIG:cartoon}b), giving rise to a Lorentz force
that curves the trajectories ${\bf r}(t)$ of quasiparticles according
to the equation of motion
$\ddot{\bf r}=(q/m)\dot{\bf r}\times{\bf B}$.
Now, Andreev reflection inverts the quasiparticle charge $q$,
mass $m$ and velocity $\dot{\bf r}$, so that in the vicinity of the
reflection point the acceleration of the outgoing hole is opposite to
that of the incoming electron.  Therefore, in a magnetic field the hole
trajectory (dashed line in Fig.~\ref{FIG:cartoon}b) no longer retraces
the electron trajectory (full line) \cite{REF:Levinson}, and vice
versa, thus allowing the motion to explore the billiard.  This raises
the possibility of chaotic motion, a possibility that we explore in
this paper.  Our primary conclusion is that although in zero ${\bf B}$
ABs are integrable regardless of their shape, integrability is
destroyed by the application of magnetic field for all but highly
regular shapes.

The study of ABs may provide an interesting link between two rapidly
developing fields: mesoscopic chaos \cite{REF:nanoexp} and mesoscopic
superconductivity \cite{REF:mesosuper}.  The prominent virtue of ABs,
viz., that they are integrable in zero ${\bf B}$ regardless of shape
and are rendered chaotic by nonzero ${\bf B}$, makes them attractive
from the experimental point of view.  By comparison, the integrability
of nanoscale CBs (e.g., in 2DEG heterostructures \cite{REF:nanoexp}) is
immensely fragile, being readily destroyed by unintentional shape
deformations or surface roughness.  Therefore, to obtain nontrivial
chaos in CBs, i.e., chaos caused by intentional choice of shape, one
must use state-of-the-art nanofabrication technology.  In contrast, an
AB prepared without any special attention to shape will be integrable,
nonintegrability of varying degree being achieved by adjusting an
external parameter, viz., ${\bf B}$ \cite{REF:flux}.

To explore qualitatively the implications of a magnetic field for the
integrability of ABs, we  focus on a planar two-dimensional AB in a
magnetic field perpendicular to the plane.  We take the dynamics to be
classical cyclotron motion with radius $R_{\rm c}$ $(\equiv mv/qB)$ of
a particle inside the billiard, supplemented by Andreev reflection at
the boundary. The case of CBs in a magnetic field has been studied
extensively \cite{REF:KNHT,REF:MBG92}, and the tangent map
\cite{REF:BERRY81,REF:MBG92} has proven to be a convenient approach.  A
tangent map is a variant of a Poincar{\'e} map, in which the state of
the system is monitored only at collisions with the boundary, the
remainder of the motion being obtained by a simple geometry.  Thus, the
problem reduces to one of following the sequence of reflection points
generated by the particle as it explores the billiard.

{\it A priori\/}, our phase space is four-dimensional: two components
of the position in the plane and two conjugate momenta.  Energy
conservation constrains the magnitude of the momentum, leaving one
freedom, which (as we are using the tangent map) we take to be the
angle $\alpha$ between the velocity ${\bf v}$ and the tangent to the
boundary (see Fig.~\ref{FIG:cartoon}c).  In addition, the fact that
reflections take place on the boundary leaves one further freedom,
which we take to be the arclength $s$ along the boundary. Then the
(continuous time) dynamics is replaced by the (discrete)  map:
$\{s,\alpha\}\rightarrow
\{s^{\prime}(s,\alpha),\alpha^{\prime}(s,\alpha)\}$,
embodying cyclotron motion followed by Andreev reflection.  For the
sake of convenience, we monitor only reflections of quasiparticles of
the same (say, electron) type. (Thus, e.g., for the sequence of
reflections  $0\to 1\to 2$ shown in Fig.~\ref{FIG:cartoon}a, only
reflections 0 and 2 are used to construct Poincar{\'e} sections.)

First, consider the case of an AB of arbitrary shape at $B=0$. In this
case, all trajectories, such as that depicted in the right billiard of
Fig.~\ref{FIG:cartoon}a, are trivially periodic, and the Poincar{\'e}
section for a given trajectory reduces to a single point, completely
determined by the initial conditions. We analyze the case of $B\neq 0$
for the example of an AB in the shape of a Bunimovich stadium
\cite{REF:BUN}, as shown in the top row of Fig.~\ref{FIG:map}.
Figure~\ref{FIG:map} shows Poincar{\'e} sections (bottom row) for a
selection of initial conditions, along with typical trajectories (top
row).  For convenience, we introduce the dimensionless magnetic field
$\beta\equiv R/R_{\rm c}\propto B$ and the tangential momentum
$p\equiv\cos\alpha$.  For the case of a weak field ($\beta=0.02$, left
column), $\{s,p\}$ space is apparently foliated by well-defined curves,
each curve corresponding to a particular initial condition.
Although it appears that the motion is integrable, when viewed at a finer
scale one sees the breakdown of foliation, as shown in the inset. Thus,
the motion is in fact weakly chaotic, as we have also confirmed by
examining the corresponding Lyapunov exponent.
In intermediate fields (for which the cyclotron radius is comparable to
the billiard size), the Poincar{\'e} sections (except those in a central
region) appear to fill a two-dimensional area of the $\{s,p\}$ plane with
disconnected points (\lq\lq dust\rq\rq), as shown for the case
$\beta=0.33$ in the middle column.  Such  behavior is commonly taken as
an indication of chaos \cite{REF:BERRY81,REF:MBG92}, and thus our
Poincar{\'e} sections suggest that indeed ABs are rendered chaotic by
the application of magnetic field. Notice the scar-like feature running
across the $\{s,p\}$ plane: it is a remnant of the foliation that
dominates in weaker fields. A single, additional,
quasi--one-dimensional Poincar{\'e} section, corresponding to an
initial condition deliberately chosen in the scar, is also shown.
In strong fields, when the cyclotron radius is much smaller than the
billiard size, particles move along skipping trajectories.  On the
scale of a typical skip, there is little distinction between motion
over straight and semi-circular segments of the boundary and,
therefore, the motion is less sensitive to the billiard shape.  As a
result, the Poincar{\'e} section exhibits a (partial) re-entrance of
integrability, i.e., the \lq\lq dust\rq\rq\ that arises in intermediate
fields is reorganized into some structures, as is seen in the bottom row
(for $\beta=10$). This structure bears a certain resemblance to that
found in weak fields, thus indicating a trend towards less chaotic
behavior. As with CBs \cite{REF:MBG92}, chaos is most pronounced for
intermediate fields, becoming less pronounced in both the weaker and
stronger field regimes.

To provide quantitative support for the suggestion of chaos inferred
from the inspection of Poincar{\'e} sections, we now turn to the
computation of Lyapunov exponents, which characterize the rate of
exponential divergence of trajectories having initial conditions nearby
in phase space. This is accomplished by investigating the stability of
the tangent map via the adaptation to ABs of the
method of Refs.~\cite{REF:BERRY81,REF:MBG92}.  Consider the situation
depicted in Fig.~\ref{FIG:cartoon}c. From the kinematics of circular
motion we have:
$\bbox{v}_{0}^{\prime}-\bbox{\omega}_{\rm c}\times\bbox{r}_{1}=
 \bbox{v}_{0}-\bbox{\omega}_{\rm c}\times\bbox{r}_{0},$
where $\bbox{\omega}_{\rm c}=q{\bf B}/m$ and $\bbox{r}_{0,1}$ are the
radius-vectors of the reflection points; other notations are defined in
Fig.~\ref{FIG:cartoon}c.  The tangent map is derived by varying this
equation with respect to $s$ and $p$, and relating the deviations of
two nearby trajectories $\delta{\bf q}$ ($\equiv\{\delta s,\delta p\}$)
before ($\delta{\bf q}_{0}$) and after ($\delta{\bf q}_{1}$) reflection
from point 1:
$\delta{\bf q}_{1}={\cal T}_{1,0}\cdot\delta{\bf q}_{0}$.
After some straightforward algebra, we find that ${\cal T}_{1,0}$  is
given by
\[
\!\!
\left(
\!\!\!
\begin{array}{lll}
\frac{R_{\rm c}\sigma\!(\chi\!)}{\rho_0\sigma\!(\alpha_1)}
-
\frac{\sigma\!\left(\alpha_0-\chi\!\right)}{\sigma\!(\alpha_1)}
&
\frac{-R_{\rm c}\sigma\!(\chi\!)}{\sigma\!(\alpha_0)\sigma\!(\alpha_1)}
\\
\frac{R_{\rm c}\sigma\!(\chi\!)}{\rho_0\rho_1}
\!+\!
\frac{\sigma\!\left(\alpha_0-\alpha_1-\chi\!\right)}{R_{\rm c}}
\!-\!
\frac{\sigma\!\left(\alpha_0-\chi\!\right)}{\rho_1}
\!-\!
\frac{\sigma\!\left(\alpha_1+\chi\!\right)}{\rho_0}
&
\frac{\sigma\!\left(\alpha_1+\chi\!\right)}{\sigma\!(\alpha_0)}
\!-\!
\frac{R_{\rm c}\sigma\!(\chi\!)}{\rho_1\sigma\!(\alpha_0)}
\\
\end{array}
\!\!\!
\right)
\]
where $\rho_{0,1}$ are the radii of curvature of the billiard boundary
at the reflection points 0 and 1, $\sigma(\phi)\equiv\sin\phi$, and
$\chi$ is defined in Fig.~\ref{FIG:cartoon}c \cite{REF:checks}.
After $N$ bounces, the separation $\delta{\bf q}_{N}$
is determined by the matrix
${\cal T}_{N,0}=\prod_{j=1}^{N} {\cal T}_{j,j-1}$.  The largest
Lyapunov exponent associated with a given trajectory is calculated
as $\lambda=\lim_{N\rightarrow\infty}\lambda_{N}$, where
\begin{equation}
\lambda_{N}=
N^{-1}
\ln
\Big\vert
\big|{\rm Tr}\,{\cal T}_{N,0}/2\big|
+\sqrt{\left({\rm Tr}\,{\cal T}_{N,0}/2\right)^{2}-1}
\Big\vert.
\label{EQ:lambda}
\end{equation}
We have calculated the Lyapunov exponents for a wide selection of
initial conditions $\left\{s_0,p_0\right\}$ and values of $B$. A
typical sequence $\lambda_N$  is shown in Fig.~\ref{FIG:lambda}.  The
convergence of $\lambda_N$ to a non-zero value as $N\rightarrow\infty$
provides evidence for the exponential divergence of nearby
trajectories, i.e., chaos.

We now turn to the issue of possible experimental realizations of ABs.
In one possible scheme, an AB  is formed by surrounding a 2DEG with a
superconducting contact \cite{REF:BJVW}.  The chaotic nature of
the motion can be diagnosed either by passing normal current through
the structure and measuring the conductance, as with nanoscale CBs
\cite{REF:nanoexp}, or by studying, e.g., via STM, correlations in
real-  and energy-space, which provide signatures of classical chaos at
the quantum level \cite{REF:GUTZ}. The primary demand on the
experimental realization of all billiards, including ABs, is that the
motion of the electrons inside the billiard be ballistic.  Our analysis
of Poincar{\'e} sections and corresponding Lyapunov exponents shows
that when $R_{\rm c}\sim L\sim R$, chaos is established after a few
bounces, and thus it  will not be masked by impurity scattering
provided that $L,R\ll\ell_{\rm e}$, where $\ell_{\rm e}$ is the elastic
mean free path of the N region. On the other hand, $B$ should not
exceed the (lower) critical field of the superconductor $B_{\rm c}$
and, hence, $R_{\rm c}>R_{\rm c}^{\rm min}=p_{\rm F}/eB_{\rm c}$,
where $p_{\rm F}$ refers to the N region.
Thus, it is sufficient to have $\ell_{\rm e}\agt R_{\rm c}^{\rm min}$.
Taking parameters for the Nb/InAs structure studied recently
(density of electrons $n_{\rm e}=9\times 10^{11}\,{\rm cm}^{-2}$,
Ref.~\cite{REF:BJVW}; $B_{\rm c}\approx 2000\,{\rm G}$) we obtain
$\ell_{\rm e}\agt 0.6\,\mu{\rm m}$, which is accessible via
current nanofabrication technologies. (E.g., $\ell_{\rm e}\approx
3.5\,\mu{\rm m}$ in Ref.~\cite{REF:HeKr}.)\thinspace\ A drawback of
the scheme described above is that the superconductor and the 2DEG are
metallurgically distinct, and thus the probability for normal
scattering at the interface is nonzero, at the expense of the Andreev
reflection, which results in billiard having a mixed AB/CB character.
This drawback can be eliminated by employing the proximity effect. The
AB is formed in a region of a superconductor where the
superconductivity has been suppressed, either due to the vicinity of a
normal metal island (see Ref.~\cite{REF:Tessmer}), or (with a type~I
superconductor) by the application of a magnetic field, which creates
domains of N phase. As such a scheme eliminates metallurgical
boundaries between N and S, reflection is of purely the Andreev type.

An interesting direction of further research would be to explore the
quantum mechanics of ABs. At least three directions are immediately
apparent:
(i)~spectral geometry and the Weyl-Kac problem;
(ii)~energy-level statistics, random matrix approaches and
universality; and
(iii)~spatial structure of quasiparticle wavefunctions.

We thank D.~K.~Campbell, S.-J.~Chang, C.~Eberlein, E.~A.~Jackson,
J.~P.~Sethna, M.~Tabor and D.~J.~Van Harlingen for useful discussions.
This work was supported in part by NSF via Grants DMR94-24511 and
DMR89-20538.


\figure{
(a)~Typical trajectories for specular (SR) and Andreev (AR)
reflection at $B=0$.  The CB (left) is formed by a single-particle
potential, the AB (right) by a pair-potential.
(b)~AR and SR in a magnetic field.
For AR, each cyclotron orbit is necessarily tangential to the
previous one, in contrast with SR.
(c)~Geometry of the tangent map.
A particle starts from 0 with velocity ${\bf v}_0$, follows the
cyclotron trajectory across the billiard, arriving at 1 with
velocity ${\bf v}_0^{\prime}$. Owing to the nature of AR,
the velocity ${\bf v}_1$ after the reflection at 1 is
$-{\bf v}_0^{\prime}$.
\label{FIG:cartoon}}
\figure{
Top row: a typical trajectory for a Bunimovich-stadium--shaped AB
$(L=R)$ for 3 values of the magnetic field:
(a)~$\beta=0.02$;
(b)~$\beta=0.33$;
(c)~$\beta=10$.
Bottom row: Poincar{\'e} sections in these fields constructed by
following the first 1000 bounces for the trajectories starting with
$\alpha_0=10^{\circ}, 20^{\circ},\ldots, 170^{\circ}$
(cf.~Fig.~\ref{FIG:cartoon}c) from random points on the perimeter of
the billiard.  Thin vertical lines on Poincar{\'e} sections separate
regions corresponding to straight (wider) and semi-circular
(narrower) segments of the billiard boundary.  In the Poincar{\'e}
sections for the weak field, flat segments result from almost-chordal
motion across a single semicircle.  ({\it Only\/} flat regions would
occur for a circular billiard.)\thinspace\ Similarly, curved regions
result from trajectories connecting any two of the four distinct
segments of the boundary.
Inset: segment (indicated by arrow) of the foliation,
magnified ($x\times\sim 10$; $y\times\sim 50$).
Ticks on the stadium boundaries mark the
points $s=0$; filled circles indicate the start of trajectories $s_0$.
We choose units in which the billiard perimeter is unity.
\label{FIG:map}}
\figure{
Lyapunov functions $\lambda_N$ [log-scale; see Eq.~(\ref{EQ:lambda})]
vs number of bounces $N$ for:
(b)~$\beta=0.33$;
(c)~$\beta=10$.
The associated trajectories are shown at the top of Fig.~\ref{FIG:map}.
For $\beta=0.02$ (not shown) we find $\lambda\approx 0.002$.
\label{FIG:lambda}}

\begin{references}
\bibitem{REF:mvbejp}
See, e.g., Ref.~\cite{REF:BERRY81} and references therein.
\bibitem{REF:BERRY81}
M.\ V.\ Berry, Eur.\ J.\ Phys.\ {\bf 2}, 91 (1981).
\bibitem{REF:AFA}
A.\ F.\ Andreev,
Zh.\ Eksp.\ Teor.\ Fiz.\
{\bf 46\/}, 1823 (1964)
[Sov.\ Phys.\ JETP {\bf 19\/}, 1228 (1964)];
Zh.\ Eksp.\ Teor.\ Fiz.\
{\bf 49\/}, 655 (1965)
[Sov.\ Phys.\ JETP {\bf 49\/}, 455  (1966)].
\bibitem{REF:quantum}
Andreev reflection is the process in which an electron-like
quasiparticle, say, moving in N impinges on an N/S interface and is
converted into a hole-like quasiparticle, which retraces the trajectory
of the incoming electron (a Cooper pair being injected into S).
Retro-reflection is not quite perfect, due to both electron-hole
interconversion and (in a magnetic field) the screening supercurrent
that circumnavigates the billiard.  However, in both cases the
violation of momentum-conservation is small (of order
$p_{\rm F}\Delta/E_{\rm F}\ll p_{\rm F}$,
where $p_{\rm F}$ is the Fermi momentum and $E_{\rm F}$ is the Fermi
energy) and should be negligible.
\bibitem{REF:Levinson}
See, e.g.,
V.\ F.\ Gantmakher and Y.\ B.\ Levinson,
{\sl Carrier scattering in metals and semiconductors\/}
(North-Holland, Amsterdam, 1990), p.~261.
\bibitem{REF:nanoexp}
For recent experiments and references to earlier work, see
C.\ M.\ Marcus et al.,
Phys.\ Rev.\ Lett.\ {\bf 69\/}, 506 (1992);
Phys.\ Rev.\ B      {\bf 48\/}, 2460 (1993).
\bibitem{REF:mesosuper}
See, e.g.,
{\sl Mesoscopic Superconductivity\/}
(Proc.\ NATO Adv.\ Res.\ Workshop),
F.\ W.\ J.\ Hekking, G.\ Sch{\"o}n, D.\ V.\ Averin (eds.),
Physica {\bf B203\/}, Nos.~3 and 4 (1994).
\bibitem{REF:flux}
In an AB surrounded entirely by superconductor, flux quantization
would restrict the allowed values of $B$.  This restriction can be
avoided by introducing a radial insulating strip into the superconductor.
\bibitem{REF:KNHT}
K.\ Nakamura, H.\ Thomas,
Phys.\ Rev.\ Lett.\ {\bf 61\/}, 247 (1988).
\bibitem{REF:MBG92}
O.\ Meplan, F.\ Brut, C.\ Gignoux,
J.\ Phys.\ {\bf A26}, 237 (1992).
\bibitem{REF:BUN}
L.\ A.\ Bunimovich, Funct.\ Anal.\ Appl.\ {\bf 8\/}, 254
(1974); Commun.\ Math.\ Phys.\ {\bf 65\/}, 259 (1979).
\bibitem{REF:checks}
One can check that ${\rm det}\,{\cal T}_{1,0}=-1$;
the tangent map is area-preserving.  Moreover,
$\lim_{R_{\rm c}\to\infty}{\cal T}_{2,1}{\cal T}_{1,0}={\openone}$,
consistent with chordal motion in zero $B$.
\bibitem{REF:BJVW}
A related structure (Nb/InAs) has recently been fabricated:
A.\ Dimoulas et al.,
Phys.\ Rev.\ Lett.\ {\bf 74\/}, 602 (1995).
\bibitem{REF:HeKr}
H.\ Kroemer et al.,
in Ref.~\cite{REF:mesosuper}, p.~298.
\bibitem{REF:GUTZ}
See, e.g., M.\ C.\ Gutzwiller,
{\sl Chaos in Classical and Quantum Mechanics\/}
(Springer-Verlag, New York, 1990).
\bibitem{REF:Tessmer}
For an STM technique particularly suitable for studying N/S
interfaces, see:
S.\  H.\ Tessmer, D.\  J.\ Van Harlingen, J.\  W.\ Lyding,
Phys.\ Rev.\ Lett.\ {\bf 70}, 3135 (1993).
\end{references}
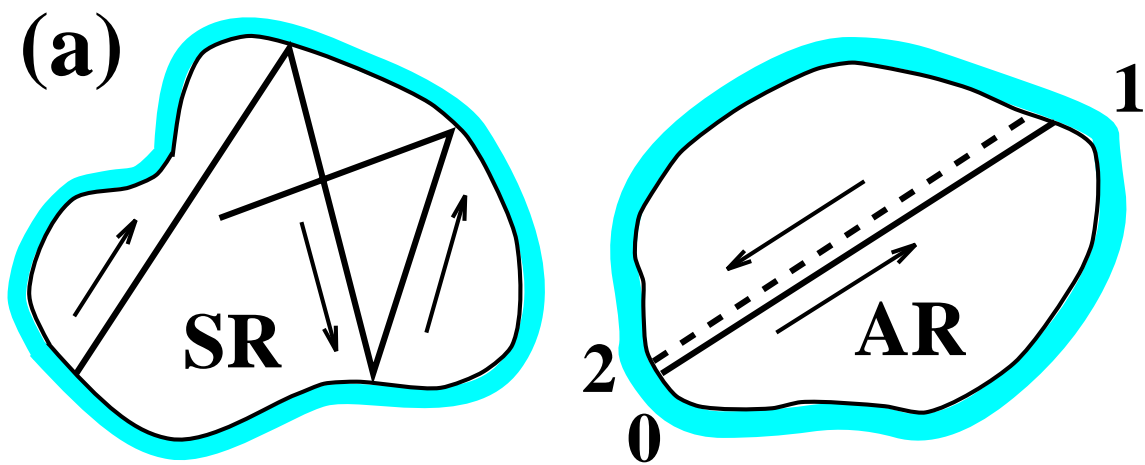
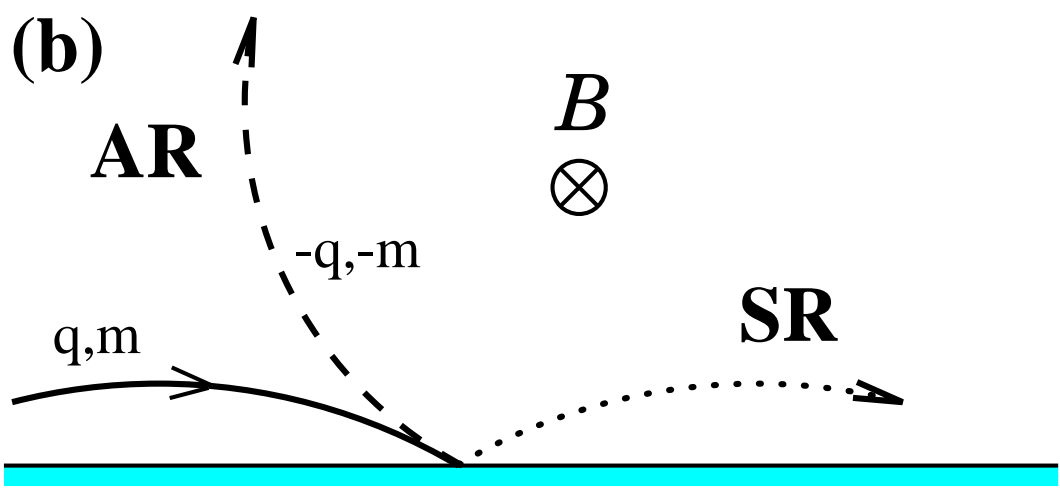
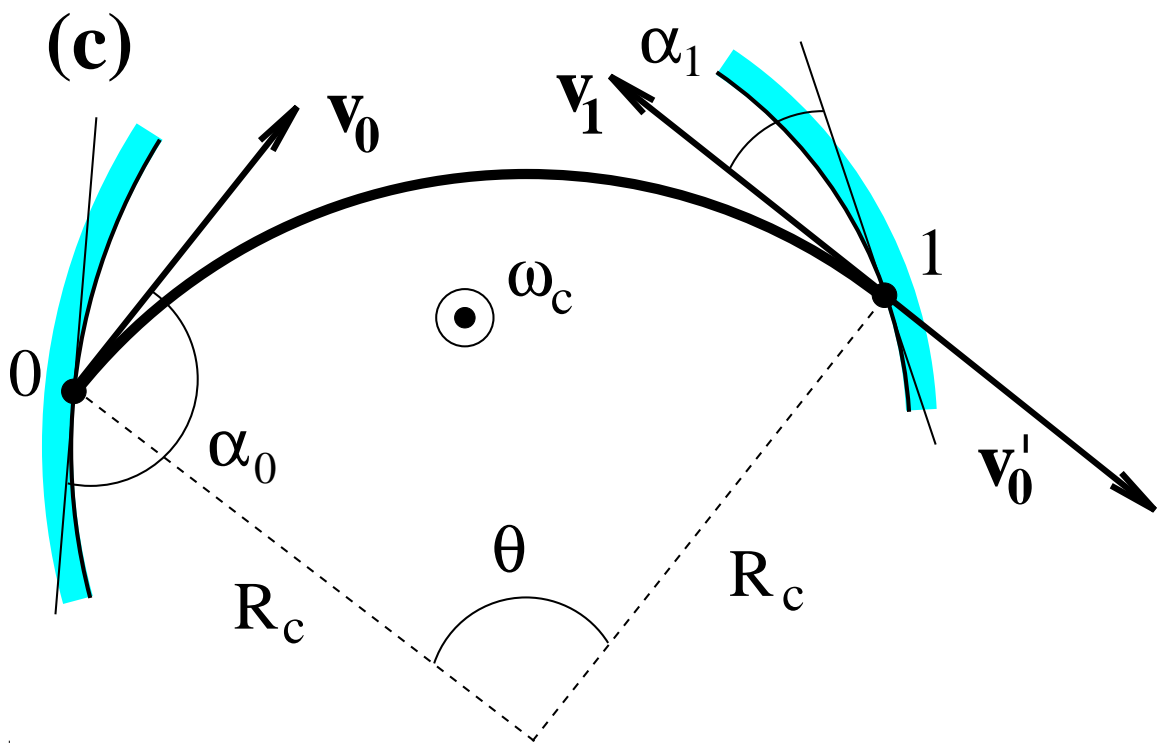
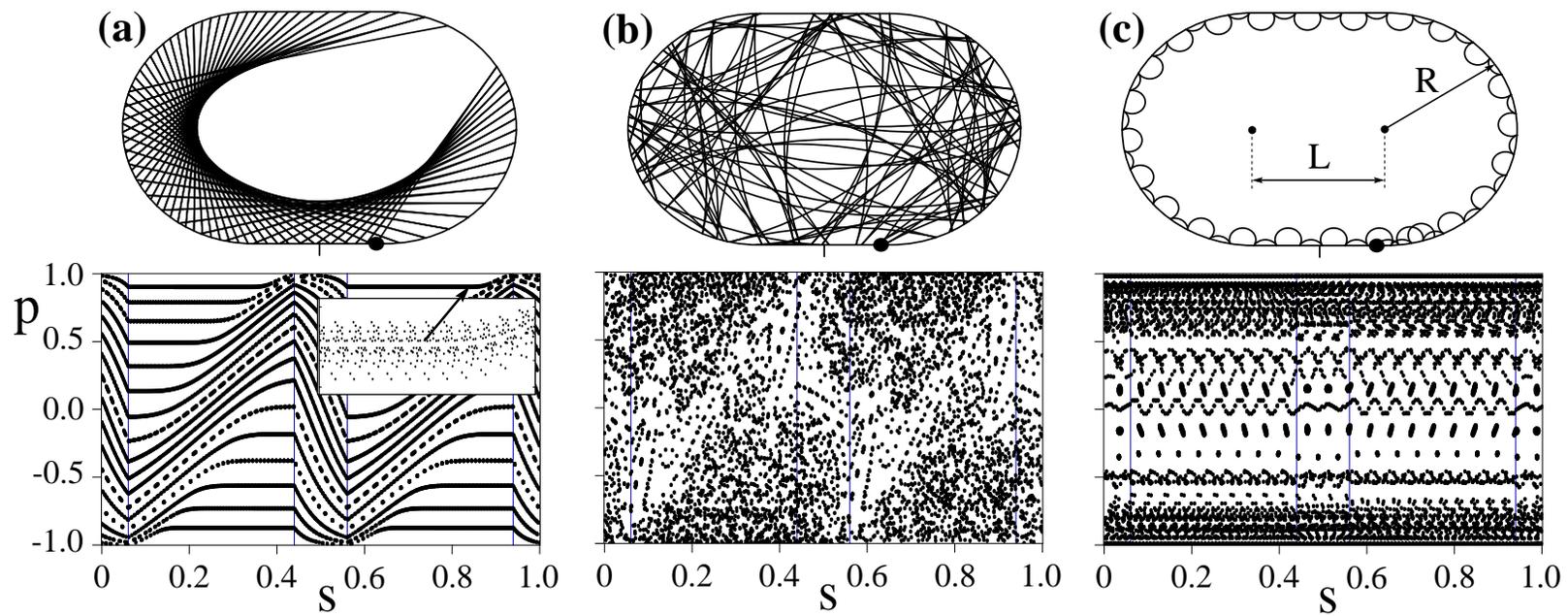
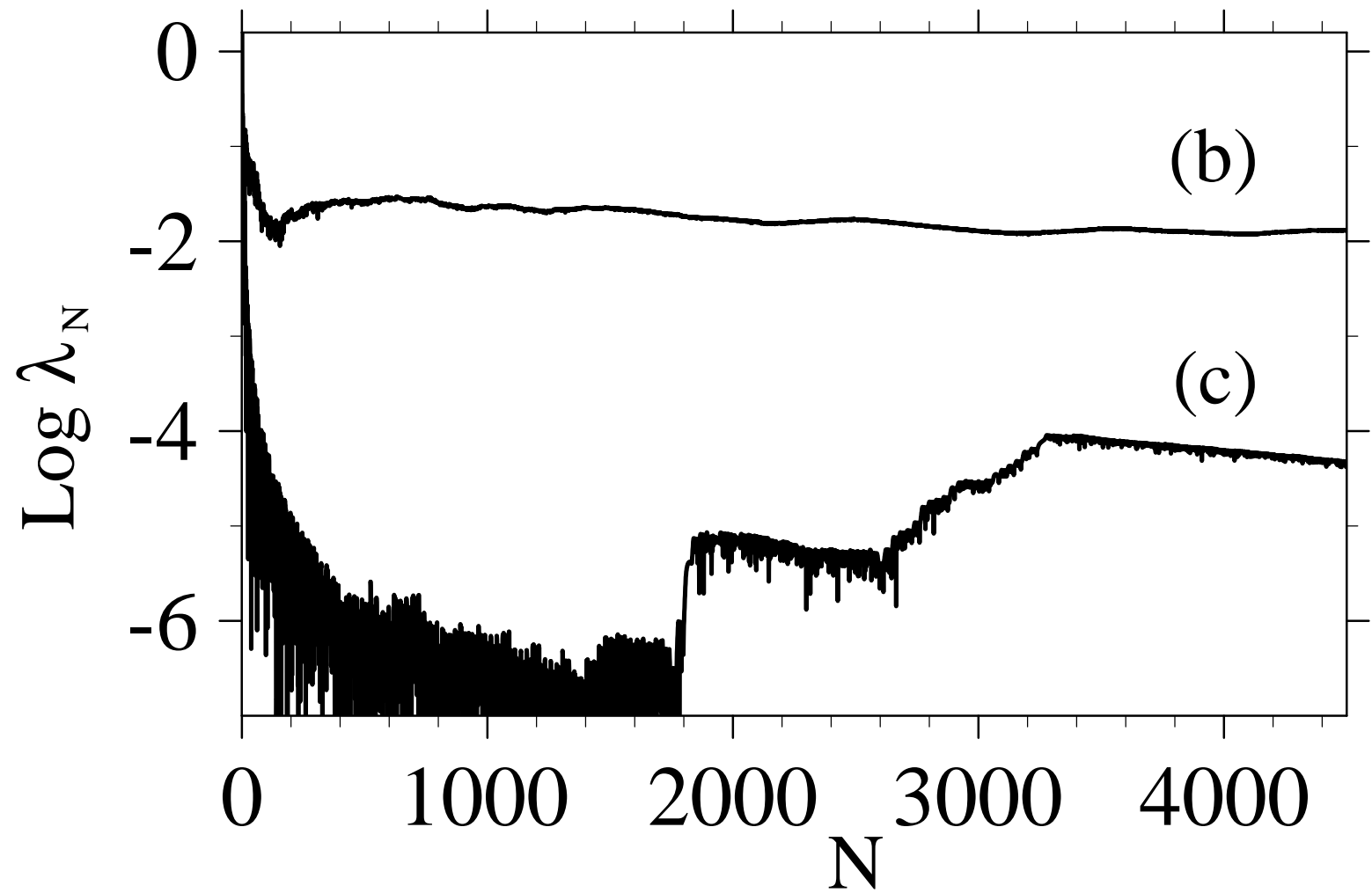
\end{document}